\documentclass[graybox]{svmult}

\usepackage{graphicx}        
\usepackage{multicol}        
\usepackage[bottom]{footmisc}

\usepackage{cite,url}
\usepackage{times}

\usepackage{amssymb}
\usepackage{amsmath}
\usepackage{amsfonts}
\usepackage{arydshln}
\usepackage{stmaryrd}

\newcommand{\Hi}{{\cal H}_\infty}
\newcommand{\R}{\mathbb{R}}
\newcommand{\C}{\mathbb{C}}
\newcommand{\UU}{\mathbf{U}}
\newcommand{\VV}{\mathbf{V}}
\newcommand{\w}{\omega}
\def\ee{{(11)}}
\def\et{{(12)}}
\def\te{{(21)}}
\def\tt{{(22)}}
\gdef \RR{{\Bbb R}}
\gdef \ZZ{{\Bbb Z}}

\begin{document}

\title*{Tuning an H-infinity controller with a given order and a structure for
interconnected systems with delays}
 \titlerunning{Tuning a fixed-order/structure $\Hi$ controller for time-delay systems}

\author{Suat Gumussoy and Wim Michiels}
\institute{Suat Gumussoy \at MathWorks, Natick MA, USA, \email{Suat.Gumussoy@mathworks.com}
\and Wim Michiels \at Department of Computer Science, KU Leuven, Belgium, \email{Wim.Michiels@cs.kuleuven.be}}

\maketitle

\abstract{
An eigenvalue based framework is developed for the $\Hi$ norm analysis and its norm minimization of coupled systems with time-delays, which are naturally described by delay differential algebraic equations (DDAEs). Fore these equations $\Hi$ norms are analyzed and their sensitivity with respect to small delay perturbations is studied. Subsequently, numerical methods for the $\Hi$ norm computation and for designing controllers minimizing the $\Hi$ norm with a prescribed structure or order, based on a direct optimization approach, are briefly addressed. The effectiveness of the approach is illustrated with a software demo. The paper concludes by pointing out the similarities with the computation and optimization of characteristic roots of DDAEs.}

\section{Introduction}
In many control applications, robust controllers are desired to achieve stability and performance requirements under model uncertainties and exogenous disturbances \cite{sg:zhou}. The design requirements are usually defined in terms of $\Hi$ norms of  closed-loop transfer functions including the plant, the controller and weights for uncertainties and disturbances. There are robust control methods to design the optimal $\Hi$ controller for linear finite dimensional multi-input-multi-output (MIMO) systems based on Riccati equations and linear matrix inequalities (LMIs), see e.g.~\cite{sg:DGKF,sg:GahinetApkarian_HinfLMI} and the references therein. The order of the controller designed by these methods is typically larger or equal to the order of the plant. This is a restrictive condition for high-order plants, since low-order controllers are desired in a practical implementation. The design of fixed-order or low-order $\Hi$ controller can be translated into a non-smooth, non-convex optimization problem.  Recently fixed-order $\Hi$ controllers have been successfully designed for finite dimensional linear-time-invariant (LTI) MIMO plants using a direct optimization approach \cite{sg:suatHIFOO}. This approach allows the user to choose the controller order and tunes the parameters of the controller to minimize the $\Hi$ norm under consideration. An extension to a class of retarded time-delay systems has been described in \cite{sg:bfgbookchapter}.

In this work we design a fixed-order or fixed-structure $\Hi$ controller in a feedback interconnection with a time-delay system. The closed-loop system is a delay differential algebraic system and its state-space representation is written as
\begin{equation}\label{sg:system}
\left\{\begin{array}{l}
E \dot x(t)= A_0 x(t)+\sum_{i=1}^m A_i x(t-\tau_i) +B w(t), \\
z(t)=C x(t).
\end{array}\right.
\end{equation}
The time-delays $\tau_i$, $i=1,\ldots,m$ are positive real numbers and the capital letters are real-valued matrices with appropriate dimensions. The input $w$ and output $z$ are disturbances and signals to be minimized to achieve design requirements and some of the system matrices include the controller parameters.

The system with the closed-loop equations (\ref{sg:system}) represents all interesting cases of the feedback interconnection of a time-delay plant and a controller. The transformation of the closed-loop system to this form can be easily done by first augmenting the system equations of the plant and controller. As we shall see, this augmented system can subsequently be brought in the form (\ref{sg:system}) by introducing slack variables to eliminate input/output delays and direct feedthrough terms in the closed-loop equations. Hence, the resulting system of the form (\ref{sg:system}) is obtained directly without complicated elimination techniques that may even not be possible in the presence of time-delays.

As we shall see, the $\Hi$ norm of DDAEs may be sensitive to arbitrarily small delay changes. Since small modeling errors are inevitable in any practical design we are interested in the smallest upper bound of the $\Hi$ norm that is insensitive to small delay changes. Inspired by the concept of strong stability of neutral equations \cite{sg:have:02}, this leads us to the introduction of the concept of  \emph{strong $\Hi$ norms} for DDAEs,  Several properties of the strong $\Hi$ norm are shown and a computational formula is obtained. The theory derived can be considered as the dual of the theory of strong stability as elaborated in \cite{sg:have:02,sg:TW-report-286,sg:Michiels:2005:NEUTRAL,sg:Michiels:2007:MULTIVARIATE} and the references therein.

 In addition, a level set algorithm for computing strong $\Hi$ norms is presented.
 Level set methods  rely on the property that
the frequencies at which a singular value of the transfer function equals a given value (the level) can be directly obtained from the solutions of a linear eigenvalue problem with Hamiltonian symmetry (see, e.g.~\cite{sg:boydbala2,sg:steinbuch,sg:byers}), allowing a two-directional search for the global maximum. For time-delay systems this eigenvalue problem is infinite-dimensional.

 Therefore, we adopt a predictor-corrector approach, where the prediction step involves a finite-dimensional approximation of the problem, and the correction serves to remove the effect of the discretization error on the numerical result. The algorithm is inspired by the algorithm for $\Hi$ computation for time-delay systems of retarded type as described in~\cite{sg:wimsimax}. However, a main difference lies in the fact that the robustness w.r.t.~small delay perturbations needs to be explicitly addressed.

The  numerical algorithm for the norm computation is subsequently applied to the design of  $\Hi$ controllers by a direct optimization approach. In the context of control of  LTI systems it is well known that  $\Hi$ norms are in general  non-convex  functions of the controller parameters which arise as elements of the closed-loop system matrices. They are typically even not everywhere smooth, although they are differentiable almost everywhere~\cite{sg:suatHIFOO}. These properties carry over to the case of strong $\Hi$ norms of DDAEs under consideration. Therefore, special optimization methods for non-smooth, non-convex problems are required. We will use a combination of BFGS, whose favorable properties in the context of non-smooth problems have been reported in \cite{sg:overtonbfgs}, bundle and gradient sampling methods, as implemented in the MATLAB code HANSO\footnote{Hybrid Algorithm for Nonsmooth Optimization, see~\cite{sg:overtonhanso}}. The overall algorithm only requires the evaluation of the objective function, i.e.,~the strong $\Hi$ norm, as well as its derivatives with respect to the controller parameters whenever it is differentiable. The computation of the derivatives is also discussed in the chapter.

The presented method is frequency domain based and builds on the eigenvalue based framework developed in~\cite{sg:bookwim}.  Time-domain methods for the $\Hi$ control of DDAEs have been described in, e.g., \cite{sg:fridman} and the references therein, based on the construction of Lyapunov-Krasovskii functionals.

\smallskip

The structure of the article is as follows. In Section \ref{sg:sec:motex} we illustrate the generality of the system description (\ref{sg:system}). The concept of asymptotic transfer function of DDAEs is introduced in Section \ref{sg:sec:tfs}. The definition and properties of the strong $\Hi$ norm of DDAEs are given in Section \ref{sg:sec:shinf}. The computation of  the strong $\Hi$ norm is described in Section \ref{sg:sec:comp_shinf}. The fixed-order $\Hi$ controller design is addressed in Section \ref{sg:sec:design}.
The concept of strong stability, fixed-order (strong) stabilization and robust stability margin optimization is summarized in Section \ref{sg:sec:stability}. Section~\ref{sg:sec:ex} is devoted to a software demo.

\subsection*{Notations} The notations are as follows. The imaginary identity is $j$. The sets of the complex, real and natural numbers are $\C, \R$, $\mathbb{N}$ respectively. The sets of nonnegative and strictly positive real numbers are $\R^+,\R_0^+$. The matrix of full column rank whose columns span the orthogonal complement of $A$ is shown as $A^{\bot}$. The zero and identity matrices are $0$ and $I$. A rectangular matrix with dimensions $n \times m$ is $A_{n\times m}$ and when square, it is abbreviated as $A_n$. The i$^\mathrm{th}$ singular value of $A$ is $\sigma_i(A)$ such that $\sigma_1(\cdot)\geq\sigma_2(\cdot)\geq \cdots$. The short notation for $(\tau_1,\ldots,\tau_m)$ is $\vec \tau\in\mathbb{R}^m$. The open ball of radius $\epsilon\in\R^+$ centered at $\vec\tau\in(\R^+)^m$ is defined as $\mathcal{B}(\vec \tau,\epsilon):=\{\vec\theta\in(\R)^m : \|\vec\theta-\vec \tau\|<\epsilon\}$.

\section{Motivating examples} \label{sg:sec:motex}

With some simple examples we illustrate the generality of the system description (\ref{sg:system}).

\begin{example} \label{sg:elim:connect}
Consider the feedback interconnection of the system and the controller as
\[
\left\{\begin{array}{lll}
\dot x(t)&=&A x(t)+B_1 u(t)+B_2w(t),\\
 y(t)&=& C x(t)+D_1 u(t),\\
 z(t)&=& F x(t),
\end{array}\quad \textrm{and}\quad  u(t)=K y(t-\tau).\right.
\]
For $\tau=0$ it is possible to eliminate the output and controller equation, which results in the closed-loop system
\begin{equation}\label{sg:elimination}
\left\{\begin{array}{lll}
\dot x(t)&=& A x(t)+B_1 K (I-D_1 K)^{-1} C x(t)+B_2 w(t), \\
z(t) & =& F x(t).
\end{array}\right.
\end{equation}
This approach is for instance taken in the software package HIFOO~\cite{sg:Burke-hifoo}.
If $\tau\neq 0$, then the elimination is not possible any more. However, if we let $X=[x^T\ u^T y^T]^T$ we can describe the system by the equations
\[
\left\{\begin{array}{l}
\left[\begin{array}{ccc}
I & 0 & 0 \\ 0 &0 & 0 \\ 0& 0&0
\end{array}\right]\dot X(t)=
\left[\begin{array}{ccc}
A & B_1 & 0 \\
C & D_1 &-I\\
0& I & 0
\end{array}\right] X(t)-
\left[\begin{array}{ccc}
0 & 0 & 0 \\
0 & 0 & 0\\
0& 0 & K
\end{array}\right] X(t-\tau) +
\left[\begin{array}{c}B_2\\0\\0
\end{array}\right] w(t),
\\
z(t)=\left[\begin{array}{cc c} F & 0&0  \end{array}\right]X(t),
 \end{array}\right.
 \] which are of the form (\ref{sg:system}).  Furthermore, the dependence of the matrices of the closed-loop system on the controller parameters, $K$, is still linear, unlike in (\ref{sg:elimination}).
\end{example}
\begin{example} \label{sg:elim:feedthru}
The presence of a direct feedthrough term from $w$ to $z$, as in
\begin{equation}\label{sg:ex2}
\left\{\begin{array}{lll}
\dot x(t)&=& Ax(t)+A_1 x(t-\tau)+B w(t),\\
z(t)&=&F x(t)+D_2 w(t),
\end{array}\right.
\end{equation}
can be avoided by introducing a slack variable. If we let $X=[x^T\ \gamma_w^T]^T$, where $\gamma_w$ is the slack variable, we can bring (\ref{sg:ex2}) in the form (\ref{sg:system}):
\[
\left\{\begin{array}{l}
\left[\begin{array}{cc}
I & 0 \\ 0 &0
\end{array}\right]\dot X(t)=
\left[\begin{array}{cc}
A & 0 \\ 0 & -I
\end{array}\right] X(t) +
\left[\begin{array}{cc}
 A_1 & 0 \\ 0 & 0
\end{array}\right] X(t-\tau) +
\left[ \begin{array}{l} B\\ I
\end{array}\right] w(t),
\\
z(t)=[F\ D_2]\ X(t).
\end{array}\right.
\]
\end{example}
\begin{example} \label{sg:elim:inputdelay}
The system
\[
\left\{\begin{array}{lll}
\dot x(t) &=& A x(t)+B_1 w(t)+B_2 w(t-\tau),\\
z(t)&=& C x(t),
\end{array}\right.
\]
can also be brought in the standard form (\ref{sg:system}) by a slack variable. Letting $X=[x^T \gamma_w^T]^T$ we can express
\[
\left\{\begin{array}{lll}
\dot X(t) &=&
\left[\begin{array}{cc}
A & B_1 \\
0 & -I
\end{array}\right]
X(t)
+
\left[\begin{array}{cc}
0 & B_2 \\
0 & 0
\end{array}\right]
X(t-\tau)
+
\left[\begin{array}{c}
0 \\ I
\end{array}\right] w(t),
\\
z(t)&=& [C\ \  0]\ X(t).
\end{array}\right.
\]
In a similar way one can deal with delays in the output $z$.
\end{example}

Using the techniques illustrated with the above examples a broad class of interconnected
systems with delays can be brought in the form (\ref{sg:system}), where the external
inputs $w$ and outputs $z$ stem from the performance specifications expressed in terms of
appropriately defined transfer functions.

The price to pay for the generality of the framework is the increase of the dimension of the system, $n$, which affects the efficiency of the numerical methods. However, this is a minor problem in most applications because the delay difference equations or algebraic constraints are related to inputs and outputs, and
the number of inputs and outputs is usually much smaller than the number of state variables.

\section{Transfer functions} \label{sg:sec:tfs}

Let $\mathrm{rank}(E)=n-\nu$, with $\nu\leq n$, and let the columns
of matrix $U\in\RR^{n\times \nu}$, respectively $V\in\RR^{n\times \nu}$, be a (minimal) basis for
the left, respectively right null space, that is, $U^T E=0$, $E V=0$.

The equations (\ref{sg:system}) can be separated into coupled delay differential and delay  difference equations. When we define $\mathbf{U}= \left[{U^{\perp}}\  U\right]$, $\mathbf{V}=\left[V^{\perp}\ V\right]$, a pre-multiplication of (\ref{sg:system}) with $\UU^T$ and the substitution $x=\VV\ [x_1^T\ x_2^T]^T$, with $x_1(t)\in\RR^{n-\nu}$ and $x_2(t)\in\RR^\nu$, yield the coupled equations
\begin{equation}\label{sg:coupled}
\left\{\begin{array}{ccl}
%
E^\ee \dot x_1(t)&=& \sum_{i=0}^m A_i^\ee x_1(t-\tau_i) +\sum_{i=0}^m A_i^\et x_2(t-\tau_i)+B_1 w(t), \\
0&=&A_0^\tt x_2(t)+ \sum_{i=1}^m A_i^\tt x_2(t-\tau_i)+\sum_{i=0}^m A_i^\te x_1(t-\tau_i)+B_2 w(t), \\
z(t)&= &C_1 x_1(t)+C_2 x_2(t),
\end{array}\right.
\end{equation}
where
\begin{eqnarray}
\nonumber A_i^\ee&=& {U^{\perp}}^T A_i V^{\perp}, \quad A_i^\et= {U^{\perp}}^T A_i V, \\
\nonumber A_i^\te&=& {U}^T A_i V^{\perp}, \quad A_i^\tt= {U}^T A_i V,\quad \textrm{for}\quad i=0,\ldots,m
\end{eqnarray}
and
\[
E^\ee= {U^{\perp}}^T E V^{\perp},\quad B_1={U^{\perp}}^T B,\quad B_2= U^T B, \quad C_1=C V^{\perp}, C_2=C. V
\]

We assume two nonrestrictive conditions:  matrix $U^T A_0 V$ is nonsingular and the zero solution of system (\ref{sg:system}), with $w\equiv0$, is strongly exponentially stable which is a necessary assumption for $\Hi$ norm optimization. For implications of the assumptions, we refer to \cite{sg:hinfdae}.

From (\ref{sg:coupled}) we can write the transfer function of the system (\ref{sg:system}) as
\begin{eqnarray}
\label{sg:T} T(\lambda)&:=&C(\lambda E-A_0-\sum_{i=1}^m A_i e^{-\lambda \tau_i})^{-1}B, \\
&=&[C_1 \ \ C_2]\left[\begin{array}{rr}\lambda E^\ee-A_{11}(\lambda) & -A_{12}(\lambda)\\
-A_{21}(\lambda) & -A_{22}(\lambda)
 \end{array}\right]^{-1} \left[\begin{array}{c}B_1\\ B_2 \end{array}\right], \label{sg:transferblock}
\end{eqnarray}
with $A_{kl}(\lambda)=\sum_{i=0}^m A_i^{(kl)} e^{-\lambda\tau_i},\ \ k,l\in\{1,2\}$.

The {\it asymptotic} transfer function of the system (\ref{sg:system}) is defined as
\begin{eqnarray}
\label{sg:Ta} T_a(\lambda) &:=&-C V (U^T A_0 V +\sum_{i=1}^m U^T A_i V e^{-\lambda\tau_i} )^{-1} U^TB \\
\nonumber &=&-C_2 A_{22}(\lambda)^{-1} B_2.
\end{eqnarray}

The terminology stems from the fact that the transfer function $T$ and the asymptotic transfer function $T_a$ converge to each other for high frequencies.

The $\Hi$ norm of the transfer function $T$ of the \emph{stable} system (\ref{sg:system}), is defined as
\[
\|T(j\w)\|_\infty:=\sup_{\w\in \R} \sigma_{1} \left( T(j\w) \right).
\]
Similarly, we can define the $\Hi$ norm of $T_a$.
\smallskip

\section{The strong H-infinity norm of time-delay systems} \label{sg:sec:shinf}

In this section we analyze continuity properties of the $\Hi$ norm of the transfer function $T$ with respect to delay perturbations, and summarize the main results of \cite{sg:hinfdae}, to which we refer for the proofs. The function
\begin{equation}\label{sg:defTdel}
\vec\tau\in(\RR_{0}^+)^{m}\mapsto \|T(j\w,\vec\tau)\|_\infty
\end{equation}
is, in general, not continuous, which is inherited from the behavior of the asymptotic transfer function, $T_a$, more precisely the function
\begin{equation}\label{sg:defTadel}
\vec\tau\in(\RR_{0}^+)^{m}\mapsto \|T_a(j\w,\vec\tau)\|_\infty.
\end{equation}
We start with a motivating example

\begin{figure}[!h]
    \begin{minipage}[t]{0.45\textwidth}
        \vspace{0pt}
        \includegraphics[width=\linewidth]{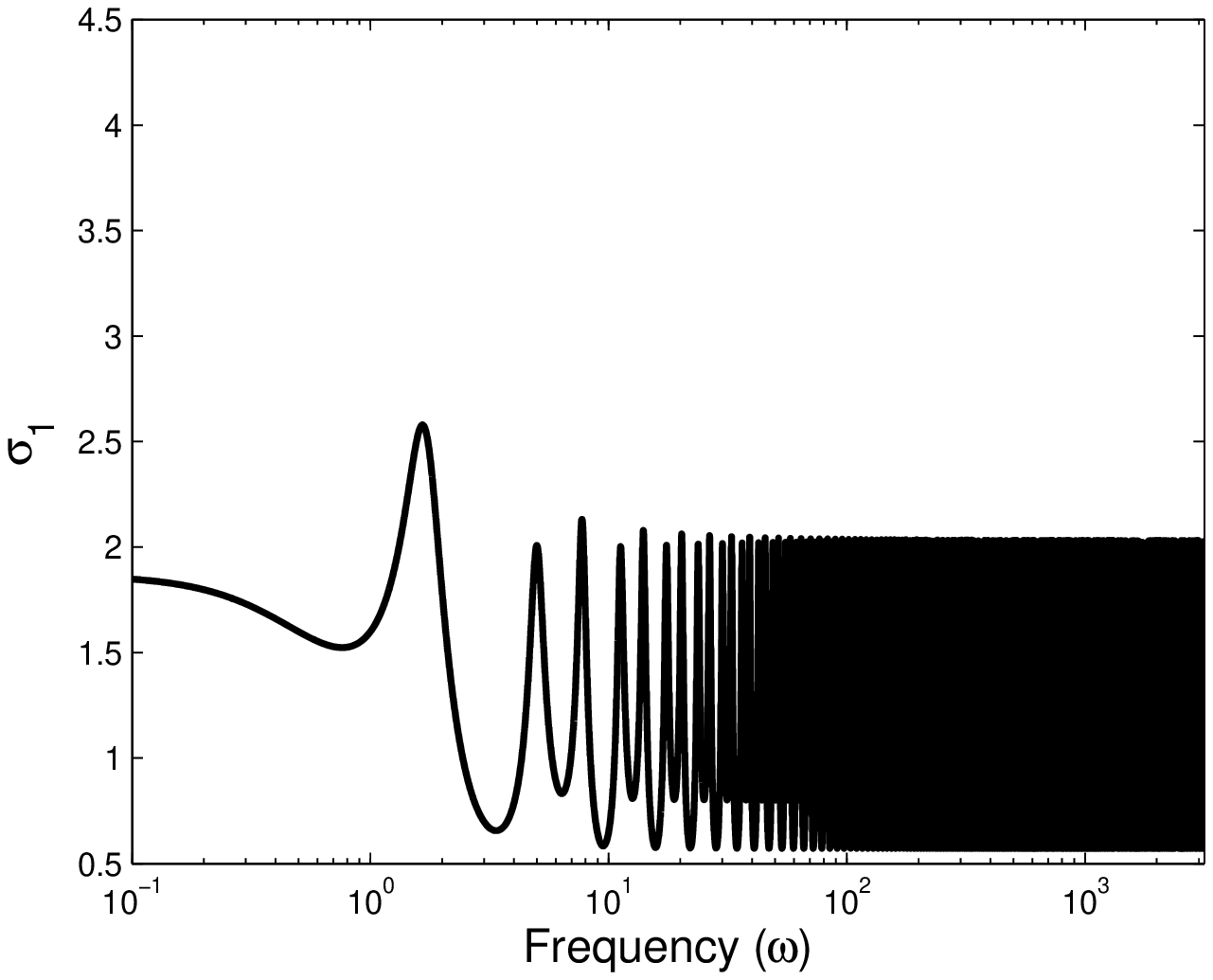}
        \caption{\label{sg:fig:svd1} The maximum singular value plot of $T(j\w,\vec\tau)$ for $(\tau_1,\tau_2)=(1,2)$ as a function of $\omega$.}
    \end{minipage}
\hfill
    \begin{minipage}[t]{0.45\textwidth}
        \vspace{0pt}\raggedright
        \includegraphics[width=\linewidth]{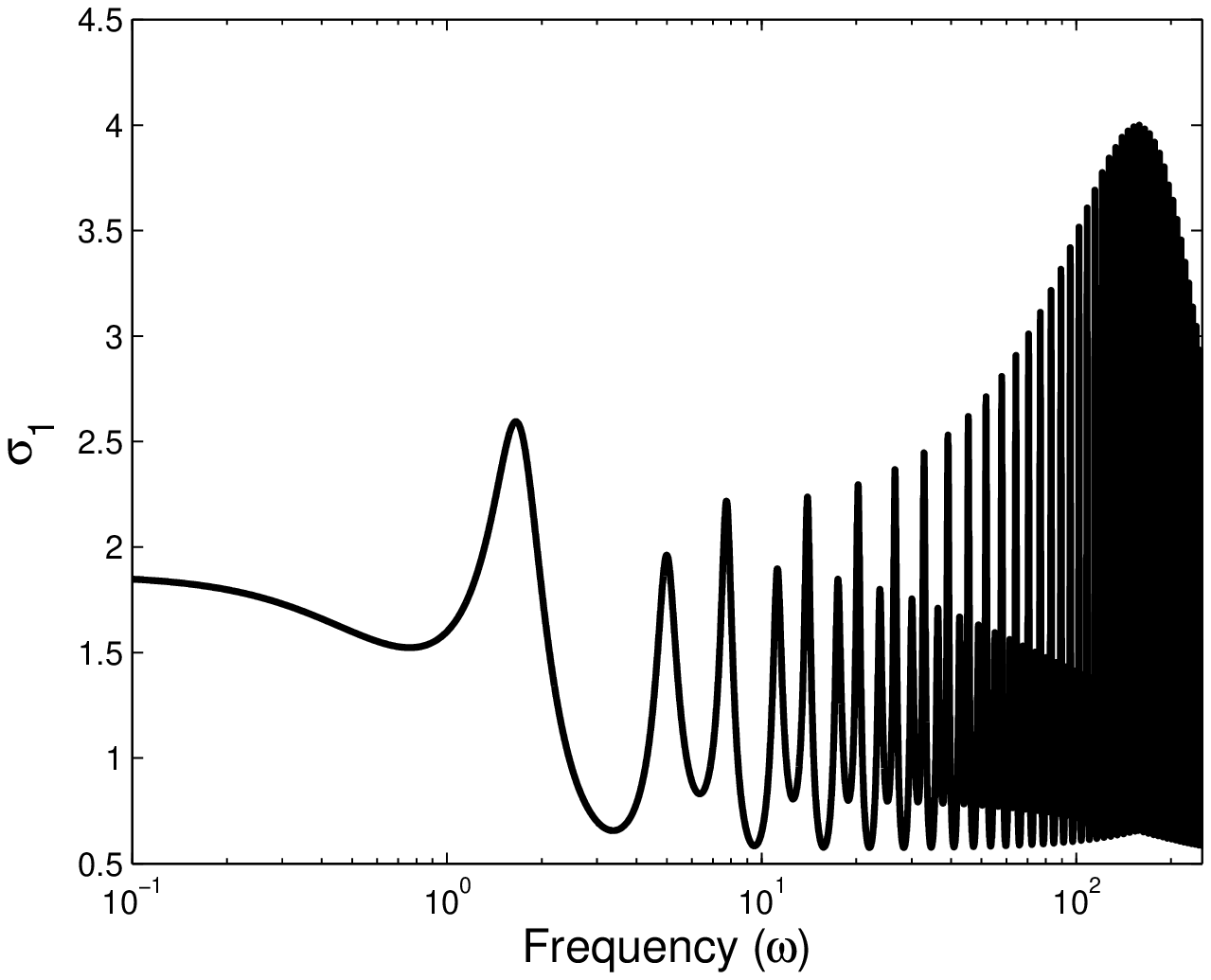}
        \caption{\label{sg:fig:svd099} The maximum singular value plot of $T(j\w,\vec\tau)$ for $(\tau_1,\tau_2)=(0.99,2)$ as a function of $\omega$.}
   \end{minipage}
\end{figure}

\begin{figure}[!h]
    \begin{minipage}[t]{0.45\textwidth}
        \vspace{0pt}\raggedright
        \includegraphics[width=\linewidth]{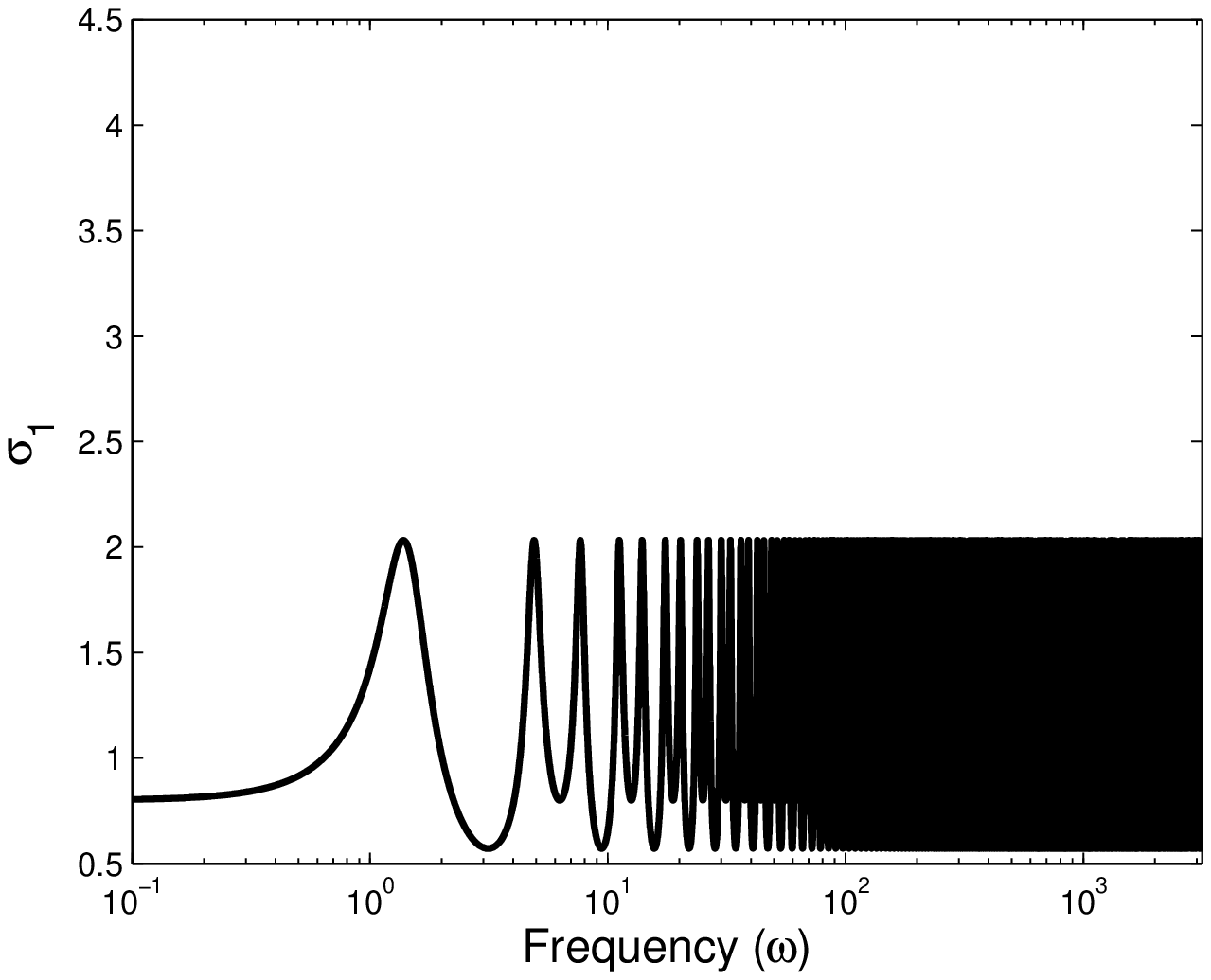}
        \caption{\label{sg:fig:svd1Ta} The maximum singular value plot of $T_a(j\w,\vec\tau)$ for $(\tau_1,\tau_2)=(1,2)$ as a function of $\omega$.}
   \end{minipage}
\hfill
    \begin{minipage}[t]{0.45\textwidth}
        \vspace{0pt}
        \includegraphics[width=\linewidth]{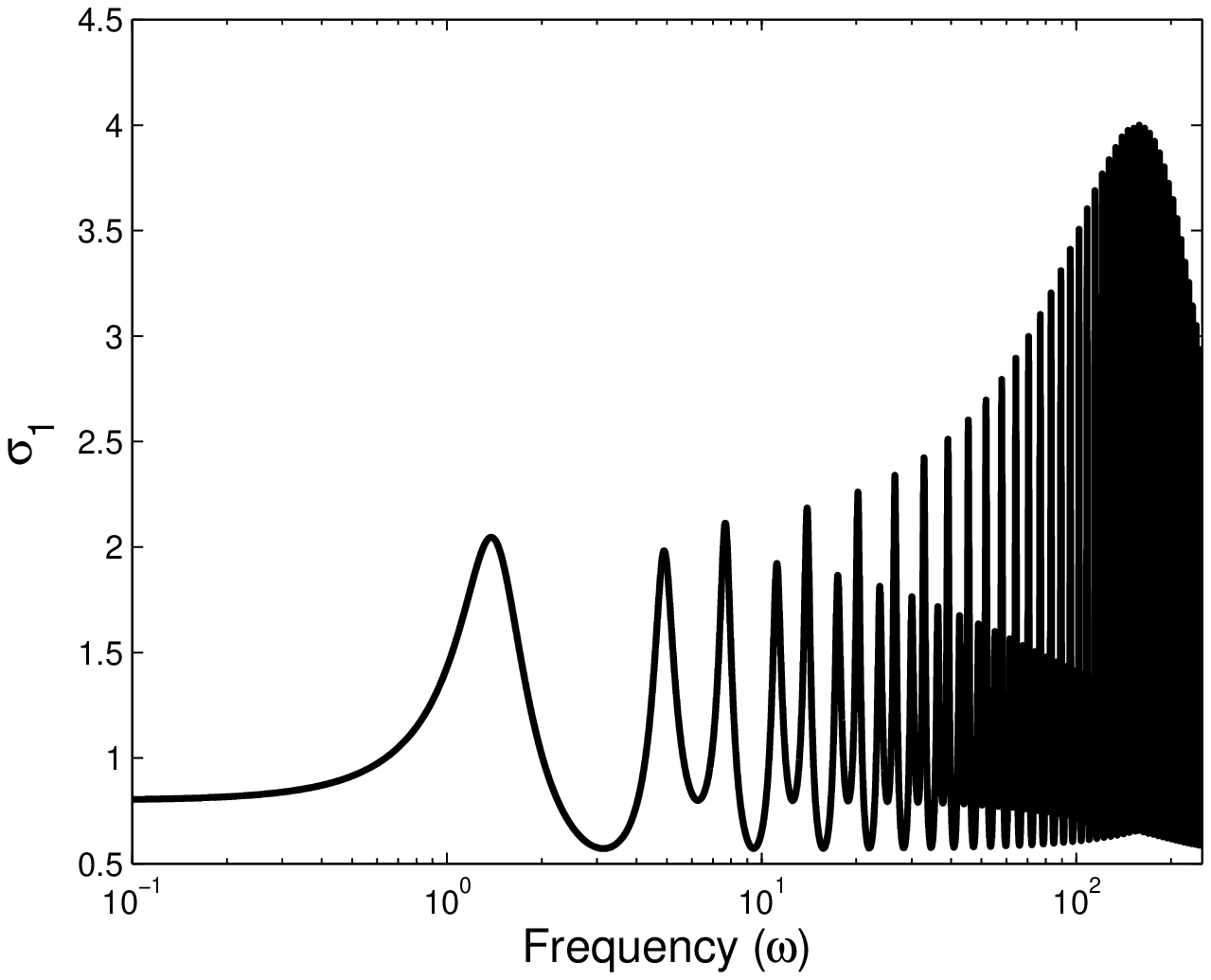}
        \caption{\label{sg:fig:svd099Ta} The maximum singular value plot of $T_a(j\w,\vec\tau)$ for $(\tau_1,\tau_2)=(0.99,2)$ as a function of $\omega$.}
        \end{minipage}
\end{figure}

\begin{example} \label{sg:ex:TandTa}
Let the transfer function $T$ be defined as
\begin{equation} \label{sg:Tex}
T(\lambda,\vec\tau)=\frac{\lambda+2.1}{(\lambda+0.1)(1-0.25e^{-\lambda\tau_1}+0.5e^{-\lambda\tau_2})+1}
\end{equation} where $(\tau_1,\tau_2)=(1,2)$. The transfer function $T$ is stable, its $\Hi$ norm is $2.5788$, achieved at $\w=1.6555$ and the maximum singular value plot is given in Figure \ref{sg:fig:svd1}. The high frequency behavior is described by the asymptotic transfer function
\begin{equation} \label{sg:Taex}
T_a(\lambda,\vec\tau)=\frac{1}{(1-0.25e^{-\lambda\tau_1}+0.5e^{-\lambda\tau_2})},
\end{equation}
whose $\Hi$ norm is equal to $2.0320$, which is less than $\|T(j\w,\vec\tau)\|_\infty$. However, when the first time delay is perturbed to $\tau_1=0.99$, the $\Hi$ norm of the transfer function $T$ is $3.9993$, reached at $\w=158.6569$, see Figure \ref{sg:fig:svd099}. The $\Hi$ norm of $T$ is quite different from that for $(\tau_1,\tau_2)=(1,2)$. A closer look at the maximum singular value plot of the asymptotic transfer function $T_a$ in Figure \ref{sg:fig:svd1Ta} and \ref{sg:fig:svd099Ta} show that the sensitivity is due to the transfer function $T_a$. Even if the first delay is perturbed slightly, the problem is not resolved, indicating that the functions (\ref{sg:defTdel}) and (\ref{sg:defTadel}) are discontinuous at $(\tau_1,\tau_2)=(1,2)$. When the delay perturbation tends to zero, the frequency where the maximum in the singular value plot of the asymptotic transfer function $T_a$ is achieved moves towards infinity.
\end{example}

The above example illustrates that the $\Hi$ norm of the transfer function $T$ may be sensitive to \emph{infinitesimal} delay changes. On the other hand, for any $\omega_{\max}>0$, the function
\[
\vec\tau\mapsto \max_{[0,\ \omega_{\max}]} \sigma_1(T(j w,\vec\tau)),
\]
where the maximum is taken over a compact set, is continuous, because a discontinuity would be in contradiction with the continuity of the maximum singular value function of a matrix. Hence, the sensitivity of the $\Hi$ norm is related to the behavior of the transfer function at high frequencies and, hence, the asymptotic transfer function $T_a$. Accordingly we start by studying the properties of the function (\ref{sg:defTadel}).

Since small modeling errors and uncertainty are inevitable in a practical design, we wish to characterize the smallest upper bound for the $\Hi$ norm of the asymptotic transfer function $T_a$ which is {\it insensitive} to small delay changes.

\begin{definition} \label{sg:def:shinfTa}
For $\vec \tau\in(\RR_{0}^+)^{m}$, let the strong $\mathcal{H}_{\infty}$ norm of $T_a$, $\interleave {T_a}(j\w,\vec \tau)\interleave_\infty$, be defined as
\[
\interleave T_a(j\w,\vec \tau)\interleave_\infty:=\lim_{\epsilon\rightarrow 0+}
\sup \{\|T_a(j\w,\vec \tau_\epsilon)\|_\infty: \vec \tau_\epsilon\in\mathcal{B}(\vec \tau,\epsilon) \cap (\RR^+)^{m} \} ,
\]
\end {definition}

Several properties of this upper bound on $\|T_a(j\w,\vec \tau)\|_\infty$ are listed below.
\begin{proposition}\label{sg:prop:Tasinfprop}
The following assertions hold:
\begin{enumerate}
\item for every $\vec\tau\in(\RR_0^+)^m$, we have
\begin{equation} \label{sg:Tasweep}
\interleave T_a(j\w,\vec \tau)\interleave_\infty=\max_{\vec\theta\in [0,\
2\pi]^m} \sigma_{1} \left( \mathbb{T}_a(\vec \theta) \right),
\end{equation}
where
\begin{equation} \label{sg:Ta_theta}
\mathbb{T}_a(\vec \theta)=-C V \left(U^T A_0 V +\sum_{i=1}^m U^T A_i V e^{-j\theta_i} \right)^{-1} U^T B;
\end{equation}
\item $\interleave T_a(j\w,\vec \tau)\interleave_\infty \geq \|T_a(j\w,\vec \tau)\|_{\infty}$ for all delays $\vec \tau$;
\item $\interleave T_a(j\w,\vec \tau)\interleave_\infty=\|T_a(j\w,\vec \tau)\|_\infty$ for rationally
independent\footnote{The $m$ components of
$\vec\tau=(\tau_1,\ldots,\tau_m)$ are rationally
independent if and only if $\sum_{k=1}^m z_k \tau_k=0,\
z_k\in\ZZ$ implies $z_k=0,\ \forall k=1,\ldots,m$. For
instance, two delays $\tau_1$ and $\tau_2$ are rationally
independent if their ratio is an irrational number.} $\vec \tau$.
\end{enumerate}
\end{proposition}

Formula (\ref{sg:Tasweep}) in Proposition \ref{sg:prop:Tasinfprop} shows that the strong $\Hi$ norm of $T_a$ is \emph{independent} of the delay values. The formula further leads to a computational scheme based on sweeping on $\vec \theta$ intervals. This approximation can be corrected by solving a set of nonlinear equations. Numerical computation details are summarized in Section~\ref{sg:sec:comp_shinf}.

\medskip

We now come back to the properties of the transfer function (\ref{sg:defTdel}) of the system (\ref{sg:system}). As we have illustrated with Example~\ref{sg:ex:TandTa}, a discontinuity of the function (\ref{sg:defTadel}) may carry over to the function (\ref{sg:defTdel}). Therefore, we define the strong $\Hi$ norm of the transfer function~$T$ in a similar way.
\begin{definition}
 For $\vec \tau\in(\RR_{0}^+)^{m}$,  the strong $\Hi$ norm of $T$, $\interleave {T}(j\w,\vec \tau)\interleave_\infty $, is given by
\[
   \interleave T(j\w,\vec \tau)\interleave_\infty:=\lim_{\epsilon\rightarrow 0+}
\sup \{\|T(j\w,\vec \tau_\epsilon)\|_\infty: \vec \tau_\epsilon\in\mathcal{B}(\vec \tau,\epsilon) \cap (\RR^+)^{m} \}.
\]
\end{definition}

The following main theorem describes the desirable property that, in contrast to the $\Hi$ norm, the strong H-infinity norm \emph{continuously} depends on the delay parameters.
It also presents an explicit expression that lays at the basis of the algorithm to compute the strong $\Hi$ norm of a transfer function, presented in the next section.
\begin{theorem}
The strong $\Hi$ norm of the transfer function of the DDAE (\ref{sg:system}) satisfies
\begin{equation} \label{sg:shinfnorm}
    \interleave T(j\w,\vec \tau)\interleave_\infty=\max\left(\|T(j\w,\vec\tau)\|_{\infty}, \interleave T_a(j\w,\vec \tau)\interleave_\infty \right),
\end{equation} where $T$ and $T_a$ are the transfer function (\ref{sg:T}) and the asymptotic transfer function (\ref{sg:Ta}).

In addition, the function
\begin{equation}\label{sg:shinfnorm2}
    \vec \tau\in(\RR^+_0)^m\mapsto \interleave T(j\w,\vec \tau)\interleave_\infty
\end{equation}
is continuous.
\end{theorem}

\smallskip

\begin{example}
We come back to Example~\ref{sg:ex:TandTa}. The $\Hi$ norm of $T$, as defined by (\ref{sg:Tex}), is $2.6422$ and the strong $\Hi$ norm of the corresponding asymptotic transfer function $T_a$ is $4$. From property (\ref{sg:shinfnorm}), we conclude that the strong $\Hi$ norm of $T$ (\ref{sg:Tex}) is $4$.
\end{example}

\begin{remark} In contrast to delay perturbations, the $\Hi$ norm of $T$ is continuous with respect to changes of the system matrices $A_i,\ldots ,A_m$, $B$ and $C$.
\end{remark}

\section{Computation of strong H-infinity norms} \label{sg:sec:comp_shinf}
We briefly outline the main steps of the strong $\Hi$ norm computation. Further details can be found in \cite{sg:hinfdae}. The algorithm for computing the strong $\Hi$ norm of the transfer function of (\ref{sg:system}) is based on property (\ref{sg:shinfnorm}). This algorithm has two important steps:
\begin{enumerate}
\item  Compute the strong $\Hi$ norm of the asymptotic transfer function $T_a$.
\item  By taking the norm in Step $1$ as the initial level set, compute the strong $\Hi$ norm of $T$ by a level set algorithm using a predictor-corrector approach.
\end{enumerate}

In the first step, the computation of $\interleave T_a(j\w,\vec \tau)\interleave_\infty$ is based on expression (\ref{sg:Tasweep}) in Proposition \ref{sg:prop:Tasinfprop}.  We obtain an approximation by restricting $\vec \theta$ in (\ref{sg:Tasweep}) to a grid,
\begin{equation} \label{sg:Taapprox}
\interleave T_a(j\w,\vec \tau)\interleave_\infty\approx\max_{\vec\theta\in\Theta_h} \sigma_{1} \left(\mathbb{T}_a(\vec \theta)\right),
\end{equation} where $\Theta_h$ is a m-dimensional grid over the hypercube $[0,\ 2\pi]^m$ and $\mathbb{T}_a(\vec \theta)$ is defined by (\ref{sg:Ta_theta}).
If a high accuracy  is required, then the approximate results may be corrected by solving a system of nonlinear equations. These equations impose that the strong $\Hi$ norm value is the maximum singular value of $\mathbb{T}_a(\vec\theta)$, and that the derivatives of this singular value with respect to the elements of $\vec \theta$ are zero.

In most practical  problems, the number of delays to be considered in $\mathbb{T}_a(\vec\theta)$ is much smaller than the number of system delays, $m$, because most of the time-delays do not appear in $\mathbb{T}_a(\vec\theta)$. This significantly reduces the computational cost of the sweeping in (\ref{sg:Taapprox}). Note that in a control application a nonzero term in (\ref{sg:Taapprox}) corresponds to a high frequency feedthrough over the control loop.

In the second step, the transfer function $T$ of (\ref{sg:system}) is approximated by a spectral discretization. The standard level set method is applied to compute an approximation of the maximum in the singular value plot and the corresponding frequency by taking as starting level  the strong $\Hi$ norm of the asymptotic transfer function $T_a$. For each level, a generalized eigenvalue problem is solved, from which intersections of singular value curves of the approximated system with the level set are computed. The predicted maxima and the frequencies are corrected by solving nonlinear equations characterizing a local maximum in the singular value plot of $T$.

\section{Fixed-order H-infinity controller design} \label{sg:sec:design}

We consider the equations
\begin{equation}\label{sg:parp}
\left\{\begin{array}{l}
E \dot x(t)=A_0(p) x(t)+\sum_{i=1}^m A_i(p) x(t-\tau_i)+Bw(t),\\
z=C x(t),
\end{array}\right.
\end{equation}
where the system matrices smoothly depend on parameters $p$.
As illustrated in Section~\ref{sg:sec:motex}, a broad class of interconnected systems can be brought into this form, where the parameters $p$ can be interpreted in terms of a parameterization of a controller. Note that, by fixing some elements of these matrices, additional structure can be imposed on the controller, e.g. a proportional-integrative-derivative (PID) like structure.

 The proposed method for designing fixed-order/ fixed-structure $\Hi$ controllers is based on a direct minimization of the strong $\Hi$ norm of the closed-loop transfer function $T$ from $w$ to $z$ as a function of the parameters $p$.  The overall optimization algorithm requires the evaluation of the objective function and its gradients with respect to the optimization parameters, whenever it is differentiable. The strong $\Hi$ norm of the transfer function $T$ can be computed as explained in the previous section. The computation of the derivatives of the norm with respect to controller parameters are given in \cite{sg:thesismarc,sg:bfgbookchapter}. The overall design procedure is fully automated and does not require any interaction with the user. Further details on the design procedure can be found in \cite{sg:hinfdae}.

\section{Strong stability, fixed-order stabilization and robust stability margin optimization} \label{sg:sec:stability}
In a practical control design, the stabilization phase is usually the first step in the overall design procedure. It is important to take the sensitivity of stability with respect to small delay perturbations into account in designing a stabilizing controller.
Similarly to the $\Hi$ norm, the spectral abscissa function, i.e., the real part of the rightmost characteristic root of a system, may namely not be a continuous function of the delays \cite{sg:TW-report-286,sg:Michiels:2007:MULTIVARIATE}. This implies that, although the characteristic roots of the overall system lie in the complex left half-plane, the system can become unstable when applying arbitrarily small delay perturbations. This discontinuity is due to the behavior of characteristic roots with high frequencies (imaginary parts).  The counterpart of the \emph{asymptotic transfer function} is the associated \emph{ delay difference equation} of the time delay system, and its characteristic roots with high imaginary parts correspond to these of the original system. \emph{The robust spectral abscissa} function introduced in \cite{sg:Michiels:2013:BookChapter} is the smallest upper bound on the spectral abscissa which continuously depends on the delays. We say that the system is \emph{strongly exponentially stable} if the exponential stability is robust with respect to small delay perturbations. A necessary and sufficient condition is given by a strictly negative  robust spectral abscissa. An algorithm to compute the robust spectral abscissa and its derivatives with respect to controller parameters is presented in \cite{sg:Michiels:2013:BookChapter}. Using this algorithm and the non-smooth, non-convex optimization methods, the robust spectral abscissa is minimized and the overall system is strongly stabilized. Note that when the standard spectral abscissa function is used as objective function, the well-known fixed-order stabilization problem is solved.

Another robustness measure is the maximum value of the spectral abscissa  when perturbations are considered to the system matrices whose Euclidean norm is bounded by a given constant $\epsilon$. This measure is called the pseudospectral abscissa and has an interpretation in terms of a $\Hi$ norm. Inherited from this connection, the pseudospectal abscissa may also be sensitive to arbitrary small delay perturbations. In accordance, the \emph{robust pseudospectal abscissa} can be defined, taking into account delay perturbations, in the same way as for the spectral abscissa and the $\Hi$ norm cases. Its computation is based on the computation of strong $\Hi$ norms. Using this computational method and non-smooth, non-convex optimization methods, the overall system can be stabilized under bounded perturbations on system matrices and arbitrary small perturbations on delays.

\section{Illustration of the software} \label{sg:sec:ex}
A MATLAB implementation of the robust stabilization algorithms is available
from
{\small
\begin{verbatim}
http://twr.cs.kuleuven.be/research/software/delay-control/.
\end{verbatim}
}
\noindent Installation instructions can be found in the corresponding README file.

We consider the following system with input delay from \cite{sg:Vanbiervliet:2008}:
\[
\dot x(t)=\verb"A" x(t)+w(t)+\verb"B" u(t-\verb"h"),\ \ \ y(t)=x(t),\ \ \ z(t)=x(t), \ \ \ u(t)=\verb"k"x(t)
\] where $\verb"h"=5$ and $\verb"k"\in\R^{1\times3}$. We start by defining the system for $w\equiv 0$:
{\small
\begin{verbatim}
A = [-0.08 -0.03 0.2;0.2 -0.04 -0.005;-0.06 0.2 -0.07];
B = [-0.1;-0.2;0.1];
C = eye(3);
p1 = tds_create({A},0,{B},5,{C},0);
\end{verbatim}
}
\noindent The uncontrolled system is unstable with a pole at $0.1081$.

In order to compute a controller , we call a routine to minimize the robust spectral abscissa with a controller order zero, \verb"nC=0",
{\small \begin{verbatim}
[k1,f1] = stabilization_max(p1,nC);
\end{verbatim}}
\noindent The controller \verb"k1" with the optimized robust spectral abscissa \verb"f1" is given by:
{\small \begin{verbatim}
k1 =

     D11: {[0.4712 0.5037 0.6023]}
    hD11: 0

f1 =

 -0.1495
\end{verbatim}} \noindent where empty fields of the controller are omitted for space considerations.

\begin{figure}
 \begin{center}
 \resizebox{5.8cm}{!}{\includegraphics{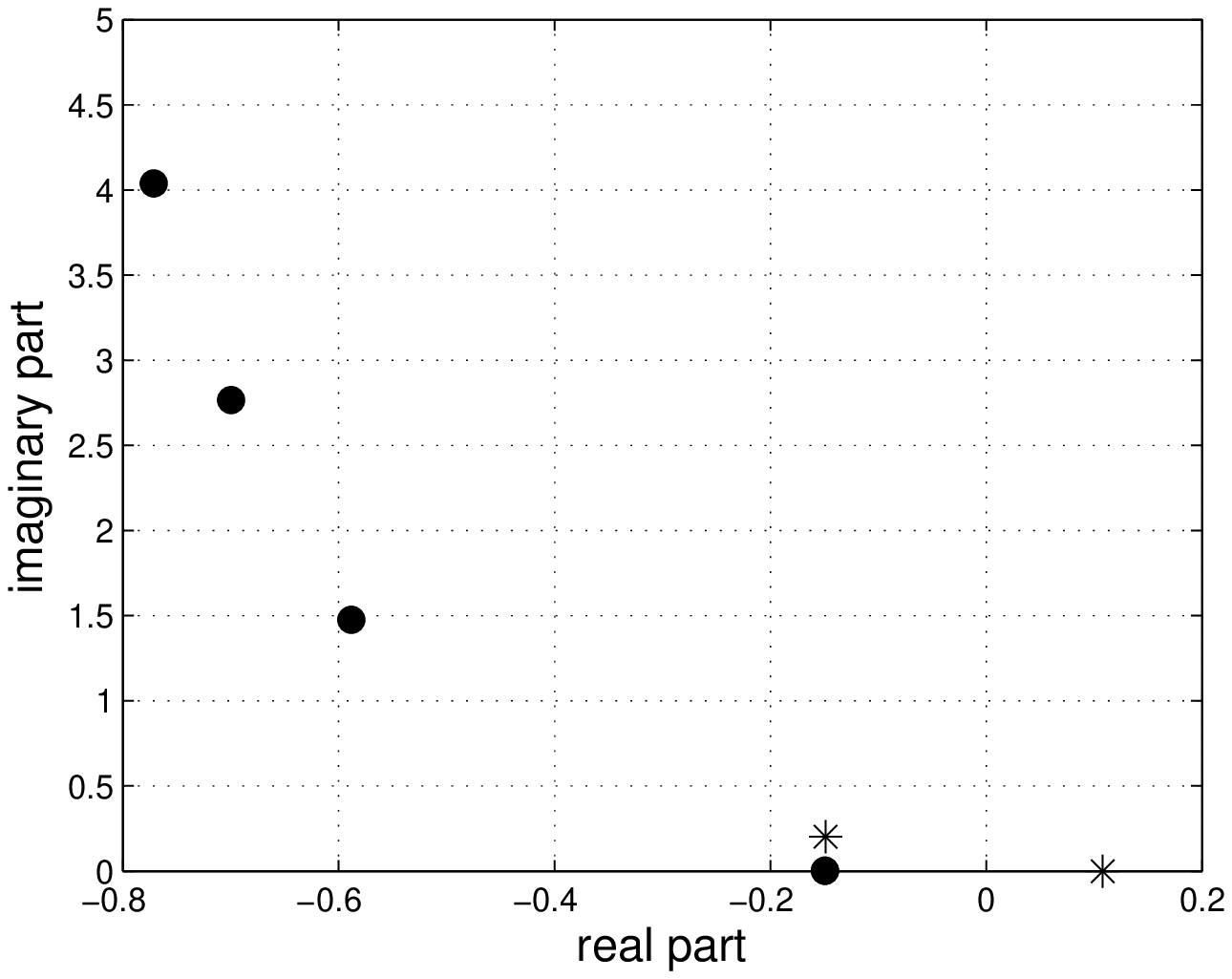}}
   \resizebox{5.8cm}{!}{\includegraphics{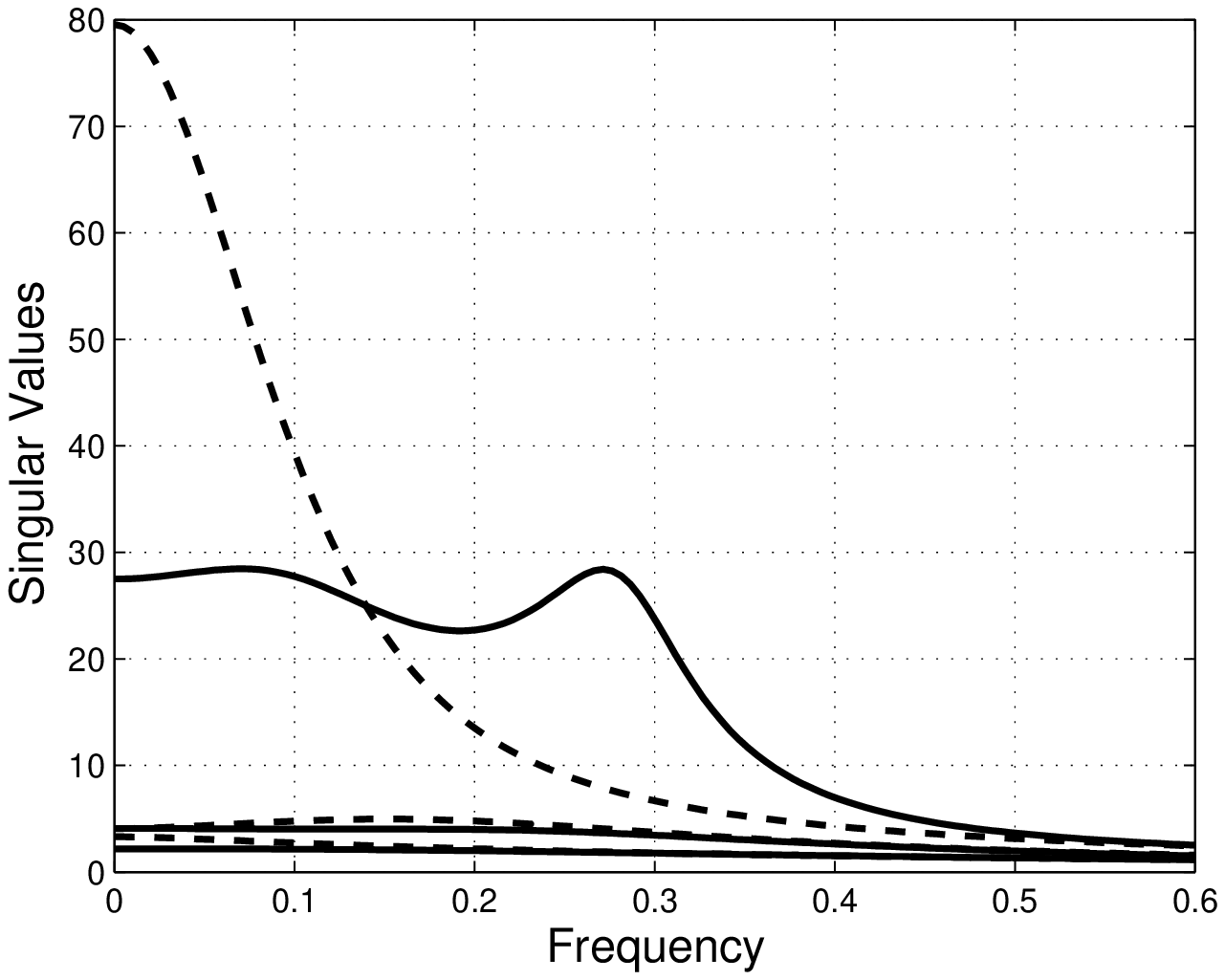}}
        \caption{ (left) Characteristic roots of the open-loop system (marked with $*$) and the closed-loop system using a static controller $k1$ (marked with $\bullet$). The closed-loop system has the rightmost characteristic root at $-0.1495$ with a multiplicity of four. (right) The singular values of the closed-loop system corresponding to the minimum of the robust spectral abscissa using a static controller $k1$ (shown in dashed lines) and corresponding to the minimum of the strong $\Hi$ norm using a static controller $k2$ (shown in straight lines). \label{sg:example}}
    \end{center}
\end{figure}

We inspect the characteristic roots of the closed-loop system with and without a controller by the following code. We first calculate the closed-loop with zero controller and the computed controller:
{\small \begin{verbatim}
k0 = tds_create({},0,{},0,{},0,{[0 0 0]},0);
clp0 = closedloop(p1,k0);
clp1 = closedloop(p1,k1);
\end{verbatim}
} We can compute all eigenvalues with real part larger than $-0.8$ for both closed-loop systems,
{\small \begin{verbatim}
options = tdsrootsoptions;
options.minimal_real_part = -0.8;
eigenvalues0 = compute_roots_DDAE(clp0,options);
eigenvalues1 = compute_roots_DDAE(clp1,options);
\end{verbatim}}
We plot the characteristic roots of the closed-loop systems,
{\small \begin{verbatim}
p0 = eigenvalues0.l1; plot(real(p0),imag(p0),’+’);
p1 = eigenvalues1.l1; plot(real(p1),imag(p1),’*’);
\end{verbatim}}
The results are displayed in Figure~\ref{sg:example} on the left. Note that the static controller stabilizes the closed-loop system by pushing the characteristic roots to the left of $s=-0.1495$ which corresponds to the computed robust spectral abscissa \verb"f1" above.

In control applications, the robustness and performance objectives are often formulated as the $\Hi$ norms of transfer functions. We can tune the controller parameters of the controller $K$ to minimize the strong $\Hi$ norm of the closed-loop system by initializing the static controller \verb"k1" computed before,

{\small \begin{verbatim}
% redefine plant with performance channels
p1 = tds_create({A},0,{eye(3)},0,{eye(3)},0,{},[],{B},5,{C});
% initialize the controller
options.K.initial = k1;
[k2,f2] = tds_hiopt(p1,nC,options);
\end{verbatim}}

\noindent The controller \verb"k2" with the optimized strong $\Hi$ norm \verb"f2" is given by:
{\small \begin{verbatim}
k2 =

     D11: {[0.7580 1.2247 0.6626]}
    hD11: 0

f2 =

 28.4167
\end{verbatim}}\noindent where empty fields of the controller are omitted for space considerations.

The singular values of the closed-loop transfer function from $w$ to $z$ are displayed in Figure~\ref{sg:example} on the right. Note that the static controller minimizing THE robust spectral abscissa has a large $\Hi$ norm, $79.5443$. This is expected since the controller is not tuned to minimize strong $\Hi$ norm but the robust spectral abscissa. The static controller minimizing strong $\Hi$ norm reduces the objective function to $28.4167$ as indicated by \verb"f2" and flattens the singular value plot as expected.\vspace{-.5cm}

\section*{Acknowledgements}\vspace{-.4cm}
{\small This article present results of the Belgian Programme on Interuniversity Poles of Attraction, initiated by the
Belgian State, Prime Minister’s Office for Science, Technology and Culture, of the Optimization in Engineering Centre OPTEC, of the project STRT1-09/33 of the K.U.Leuven Research Council and of the Project G.0712.11 of the Fund for Scientific Research -Flanders.}


\vspace{-.5cm}
\begin{thebibliography}{99.}

\bibitem{sg:boydbala2}
S.~Boyd, V.~Balakrishnan, and P.~Kabamba.
\newblock A bisection method for computing the $\mathcal{H}_{\infty}$ norm of a
  transfer matrix and related problems.
\newblock {\em Mathematics of Control, Signals and Systems}, 2:207--219, 1989.

\bibitem{sg:steinbuch}
N.A. Bruinsma and M.~Steinbuch.
\newblock A fast algorithm to compute the $\mathcal{H}_{\infty}$-norm of a
  transfer function matrix.
\newblock {\em Systems and Control Letters}, 14:287--293, 1990.

\bibitem{sg:Burke-hifoo}
J.~V. Burke, D.~Henrion, A.~S. Lewis, and M.~L. Overton.
\newblock {HIFOO} - a {\sc matlab} package for fixed-order controller design
  and {H}-infinity optimization.
\newblock In {\em Proceedings of the 5th IFAC Symposium on Robust Control
  Design}, Toulouse, France, 2006.

\bibitem{sg:byers}
R.~Byers.
\newblock A bisection method for measuring the distance of a stable matrix to
  the unstable matrices.
\newblock {\em SIAM Journal on Scientific and Statistical Computing},
  9(9):875--881, 1988.

\bibitem{sg:DGKF}
J.C. Doyle, K.~Glover, Khargonekar P.P., and Francis B.A.
\newblock State-space solutions to standard $\mathcal{H}^2$ and
  $\mathcal{H}^\infty$ control problems.
\newblock {\em IEEE Transactions on Automatic Control}, 34(8):831--847, 1989.


\bibitem{sg:fridman}
E.~Fridman and U.~Shaked.
\newblock ${H_{\infty}}$-control of linear state-delay descriptor systems: an
  {LMI} approach.
\newblock {\em Linear Algebra and its Applications}, 351-352:271--302, 2002.

\bibitem{sg:GahinetApkarian_HinfLMI}
P.~Gahinet and P.~Apkarian.
\newblock A linear matrix inequality approach to $\mathcal{H}_\infty$ control.
\newblock {\em International Journal of Robust and Nonlinear Control},
  4(4):421--448, 1994.

\bibitem{sg:hinfdae}
S.~Gumussoy and W.~Michiels.
\newblock Fixed-Order H-infinity Control for Interconnected Systems using Delay Differential Algebraic Equations.
\newblock {\em SIAM Journal on Control and Optimization}, 49(2):2212--2238, 2011.

\bibitem{sg:bfgbookchapter}
S.~Gumussoy and W.~Michiels.
\newblock Fixed-order {H}-infinity optimization of time-delay systems.
\newblock In M.~Diehl, F.~Glineur, E.~Jarlebring, and W.~Michiels, editors,
  {\em Recent Advances in Optimization and its Applications in Engineering}.
  Springer, 2010.

\bibitem{sg:suatHIFOO}
S.~Gumussoy and M.L. Overton.
\newblock Fixed-order {H}-infinity controller design via {HIFOO}, a specialized
  nonsmooth optimization package.
\newblock In {\em Proceedings of the American Control Conference}, pages
  2750--2754, Seattle, USA, 2008.

\bibitem{sg:have:02}
J.K. Hale and S.M Verduyn~Lunel.
\newblock Strong stabilization of neutral functional differential equations.
\newblock {\em IMA Journal of Mathematical Control and Information}, 19:5--23,
  2002.

\bibitem{sg:overtonbfgs}
A.~Lewis and M.L. Overton.
\newblock Nonsmooth optimization via {BFGS}.
\newblock Available from \verb|http://cs.nyu.edu/overton/papers.html|, 2009.

\bibitem{sg:TW-report-286}
W.~Michiels, K.~Engelborghs, D.~Roose, and D.~Dochain.
\newblock Sensitivity to infinitesimal delays in neutral equations.
\newblock {\em {SIAM} Journal on Control and Optimization}, 40(4):1134--1158,
  2002.

\bibitem{sg:wimsimax}
W.~Michiels and S.~Gumussoy.
\newblock Characterization and computation of {H}-infinity norms of time-delay
  systems.
\newblock {\em {SIAM} Journal on Matrix Analysis and Applications},
  31(4):2093--2115, 2010.

\bibitem{sg:bookwim}
W.~Michiels and S.-I. Niculescu.
\newblock {\em Stability and stabilization of time-delay systems. An eigenvalue
  based approach}.
\newblock SIAM, 2007.

\bibitem{sg:Michiels:2005:NEUTRAL}
W.~Michiels and T.~Vyhl{\'\i}dal.
\newblock An eigenvalue based approach for the stabilization of linear
  time-delay systems of neutral type.
\newblock {\em Automatica}, 41(6):991--998, 2005.

\bibitem{sg:Michiels:2007:MULTIVARIATE}
W.~Michiels, T.~Vyhl{\'\i}dal, P.~Z{\'\i}tek, H.~Nijmeijer, and D.~Henrion.
\newblock Strong stability of neutral equations with an arbitrary delay
  dependency structure.
\newblock {\em {SIAM} Journal on Control and Optimization}, 48(2):763--786,
  2009.

\bibitem{sg:Michiels:2013:BookChapter}
W.~Michiels, and S.~Gumussoy.
\newblock Eigenvalue based algorithms and software for the design of fixed-order stabilizing controllers for interconnected systems withtime-delays.
\newblock {\em Lecture Notes in Delays and Dynamics}, Springer (To Appear),
  2013.

\bibitem{sg:thesismarc}
{Millstone, M.}
\newblock {HIFOO} 1.5: Structured control of linear systems with a non-trivial
  feedthrough.
\newblock Master's thesis, New York University, 2006.

\bibitem{sg:overtonhanso}
M.~Overton.
\newblock {HANSO}: a hybrid algorithm for nonsmooth optimization.
\newblock Available from \verb|http://cs.nyu.edu/overton/software/hanso/|,
  2009.

\bibitem{sg:Vanbiervliet:2008}
J.~Vanbiervliet, B.~Vandereycken, W.~Michiels, and S.~Vandewalle.
\newblock A nonsmooth optimization approach for the stabilization of time-delay systems.
\newblock {\em {ESAIM} Control, Optimisation and Calculus of Variations}, 14(3):478--493,
  2008.

\bibitem{sg:zhou}
K.~Zhou, J.C. Doyle, and K.~Glover.
\newblock {\em Robust and optimal control}.
\newblock Prentice Hall, 1995.

\end{thebibliography}
\end{document}